\documentclass[aps,prb,twocolumn,floats,epsfig,pdflatex]{revtex4}
\usepackage{amsmath}
\usepackage{amsfonts}
\usepackage{graphicx}
\usepackage{amssymb}
\usepackage{amsbsy}
\usepackage{color}
\usepackage{ulem}

\newcommand{\beq}{\begin{equation}}
\newcommand{\eeq}{\end{equation}}
\newcommand{\bea}{\begin{eqnarray}}
\newcommand{\eea}{\end{eqnarray}}

\begin{document}

\title{Prethermal fragmentation in a periodically driven Fermionic chain}
\author{Somsubhra Ghosh$^1$, Indranil Paul$^2$, and
K. Sengupta$^1$}
\affiliation{$^1$School of Physical Sciences, Indian Association for the
Cultivation of Science, Kolkata 700032, India \\
$^2$Universit\'{e} Paris Cit\'{e}, CNRS, Laboratoire Mat\'{e}riaux
et Ph\'{e}nom\`{e}nes Quantiques, 75205 Paris, France}
\date{\today}

\begin{abstract}

We study a Fermionic chain with nearest-neighbor hopping and
density-density interactions, where the nearest-neighbor interaction
term is driven periodically. We show that such a driven chain
exhibits prethermal strong Hilbert space fragmentation (HSF) in the
high drive amplitude regime at specific drive frequencies
$\omega_m^{\ast}$. This constitutes the first realization of HSF for
out-of-equilibrium systems. We obtain analytic expressions of
$\omega_m^{\ast}$ using a Floquet perturbation theory and provide
exact numerical computation of entanglement entropy, equal-time
correlation functions, and the density autocorrelation of Fermions
for finite chains. All of these quantities indicate clear signatures
of strong HSF. We study the fate of the HSF as one tunes away from
$\omega_m^{\ast}$ and discuss the extent of the prethermal regime as
a function of the drive amplitude.

\end{abstract}


\maketitle

Eigenstate thermalization hypothesis (ETH) conjectures that
mid-spectrum eigenstates of an isolated quantum system are locally
thermal \cite{eth1,eth2,rev1,rev2}. A similar thermal characteristic
leading to an infinite temperature steady state is seen for
periodically driven systems \cite{fleth1,rev3}. In both cases, ETH
has been immensely successful in predicting the long-time dynamical
behavior of local operators of isolated quantum systems. However,
for periodically driven systems experimentally relevant time scales
may be significantly shorter than their thermalization times
\cite{scaling3,mblexp,tc2}. In such cases, the prethermal
characteristics of the driven system becomes relevant. It is
well-known that these prethermal phases may exhibit interesting
phenomena that have no equilibrium analogue \cite{adas1,
adas2,tista1,dsen1,luitz1,roopayan1,tc1,tc2,flscar1,flscar2}.


The violation of ETH in a quantum system comes primarily from the
loss of ergodicity. This can occur in integrable quantum systems
\cite{rev1,rev2} or in systems with strong disorder in their many
body localized phases \cite{mbl1,mblrev,mblexp}. In addition, a weak
violation of ETH can also occur for many-body Hamiltonians hosting 
quantum scars \cite{scarref1,scarref2,scarref3,scarref4,scarrev}
which may lead to long-time prethermal oscillatory dynamics, instead
of ETH predicted rapid thermalization, of local correlation
functions for certain initial states. The effect of the presence of
such scars in the Floquet Hamiltonian of periodically driven systems
has also been studied
\cite{flscar1,flscar2,flscar3,flscar4,flscar5}.

Another route to ETH violation in a quantum system occurs from
fragmentation of its Hilbert space due to the presence of kinetic
constraints on the dynamics of its constituent particles or spins
\cite{hsf1,hsf2,hsf3,hsf4,hsf5,hsf6,hsf7, hsf8,hsf9,hsf10}. The
Hamiltonian of such a system, in the computational basis, breaks
down into several dynamically disconnected blocks. For strong
Hilbert space fragmentation (HSF), the number of such blocks
increases exponentially with the system size; this is in sharp
contrast to algebraic scaling of number of disconnected symmetry
sectors.
All the studies of strong HSF, so far, have involved equilibrium
Hamiltonians; to the best our knowledge, there has been no example
of the existence of strong HSF in out-of-equilibrium quantum
systems. In this work we provide the first example of prethermal HSF
in a periodically driven Fermi chain with large drive amplitude at
special drive frequencies.

To this end, we consider a Fermion chain with a Hamiltonian given by
$H(t)= H_0(t) +H_1$
\begin{eqnarray}
&& H_0(t) =  V(t) \sum_{j=1..L} \hat n_j \hat n_{j+1} \label{fermham} \\
&& H_1 = \sum_{j=1..L} -J (c_j^{\dagger} c_{j+1} + {\rm h.c.}) +
\hat n_j (V_0 \hat n_{j+1} + V_2 \hat n_{j+2})\nonumber
\end{eqnarray}
where $c_j$ denotes the Fermion annihilation operator for the site
$j$ of the chain, $\hat n_j= c_j^{\dagger} c_j$ is the Fermion
density operator, $V_0+V(t)$ and $V_2$ are the strengths of nearest-
and next-nearest neighbor interactions respectively. In what
follows, we drive this Fermion chain by making $V(t)$ a periodic
function of time characterized by an amplitude $V_1 \gg V_0,J, V_2$
and frequency $\omega_D = 2\pi/T$, where $T$ is the time period of
the drive. The precise form of $V(t)$ depends on the drive protocol;
in this letter we shall study both continuous cosine ($V(t)= V_1
\cos \omega_D t$) and discrete square-pulse ($V(t)= -(+) V_1$ for $t
\le (>) T/2$) protocols.

The results obtained from the study of such a Hamiltonian are as
follows. First, we derive the Floquet Hamiltonian of the driven
chain in the high amplitude limit using Floquet perturbation theory
(FPT). We show that at special drive frequencies
$\omega_D=\omega_m^{\ast}$, whose analytic expression we provide,
the first order Floquet Hamiltonian, $H_F^{(1)}$, of the system
reduces to a Fermionic model with constrained hopping which is known
to exhibit HSF \cite{hsf4,hsf5}. In the high-drive amplitude regime,
the higher order correction to $H_F^{(1)}$ are small; consequently,
the driven system exhibits signatures of fragmentation over a long
prethermal regime. Second, using exact diagonalization (ED), we
compute exact evolution of the half-chain von-Neumann entanglement
entropy, $S(nT)$, starting from a random Fock state, as a function
of the number of drive cycles $n$. We find that away from
$\omega_m^{\ast}$, $S(nT)$ reaches the symmetry resolved Page value,
$S_p$, as expected for non-integrable ergodic driven systems
\cite{page1}; in contrast, at $\omega_D=\omega_m^{\ast}$ and for
large $V_1$, $S(nT)$ saturates to the Page value, $S_p^f$, of the
Hilbert space fragment of $H_F^{(1)}$ to which the initial state
belongs for a long range of $n$. Third, we compute the Fermion
density autocorrelation function
\begin{eqnarray}
C_j(nT)= \langle \psi_0 |(\hat n_j(nT)-1/2) (\hat n_j(0)-1/2)|\psi_0
\rangle \label{acor}
\end{eqnarray}
starting from a random infinite-temperature initial state
$|\psi_0\rangle$. We find that $C_j(nT)$ does not attain its ETH
predicted value at $\omega_D=\omega_m^{\ast}$ beyond a critical
$V_1$; instead it saturates to a finite value, larger than a lower
bound that can be obtained using Mazur's inequality
\cite{mazur1,hsf2,hsf3, supp1}, as expected in a system with strong
HSF \cite{hsf1}. In contrast, for $\omega_D \ne \omega_m^{\ast}$,
$C_j$ obeys ETH. Fourth, we study the equal-time correlation
function
\begin{eqnarray}
\chi_j(nT) =\langle \psi_f(nT)|\hat n_j \hat n_{j+2}
|\psi_f(nT)\rangle  \label{etcorr}
\end{eqnarray}
of Fermions starting from frozen states (Fock states
$|\psi_f\rangle$ which are eigenstates of $H_F^{(1)}$). The dynamics
of these states, at $\omega_D=\omega_m^{\ast}$, arise solely due to
presence of higher order terms in the Floquet Hamiltonian;
consequently, for large $V_1$, $\chi(nT)$ remain close to their
initial values at $\omega_D= \omega_m^{\ast}$ over a large range of
$n$ for random frozen initial states. In contrast, they saturate to
the ETH predicted value at other frequencies. Moreover, for the
frozen state $|\psi_f\rangle = |Z_2\rangle \equiv |0,1,0,1
....\rangle$, we find the existence of novel oscillatory dynamics of
$\chi_j(nT)$ when $J/V_0 \ll 1$; we provide a semi-analytic
explanation of this dynamics and tie the existence of such
oscillations to both the broken $Z_2$ symmetry of the initial Fock
state and the presence of HSF which confines the dynamics to a small
class of states in the Hilbert space.

{\it FPT and Floquet Hamiltonian}: To derive the Floquet Hamiltonian
for the driven chain, we start from Eq.\ \ref{fermham} and adapt a
perturbative scheme (FPT) where $J/V_1$ is the small parameter; this
distinguishes it from standard high-frequency Magnus expansion
\cite{rev8,dsen2,tb1}. Within this scheme, one computes the
evolution operator corresponding to $H_0(t)$ exactly; $ U_0(t,0) =
{\mathcal T} \exp[- i \int_0^t dt' H_0(t')/\hbar]$, where ${\mathcal
T}$ is the time ordering operator. The contribution of $H_1$ to the
full evolution operator $U(t,0)$ is then computed using standard
perturbation theory. To first order in perturbation, this leads to
$U_1(T,0)= (-i/\hbar) \int_0^T dt (U_0(t,0))^{\dagger} H_1 U_0(t,0)$
and yields the leading order Floquet Hamiltonian $H_F^{(1)} = i\hbar
U_1(T,0)/T$. A straightforward calculation detailed in Ref.\
\onlinecite{supp1} leads to
\begin{eqnarray}
H_F^{(1)} &=& \sum_{j=1..L} \hat n_j ( V_0 \hat n_{j+1} +V_2 \hat
n_{j+2})
\label{fl1} \\
&& -J \sum_{j=1..L} [ (1-\hat A_j^2) + f(\gamma_0) \hat A_j^2]
c_j^{\dagger} c_{j+1} +{\rm h.c.} \nonumber
\end{eqnarray}
where $\hat A_j =(\hat n_{j+2}- \hat n_{j-1})$, $\gamma_0= V_1
T/(4\hbar)$ and $f(\gamma_0) = J_0[2 \gamma_0/\pi]$ for the cosine
protocol and $f(\gamma_0)= \gamma_0^{-1} \sin \gamma_0 \exp[i
\gamma_0 \hat A_j]$ for the square-pulse protocol, where $J_0$ is
the zeroth order Bessel function.

Eq.\ \ref{fl1} represents the central result of this work; it shows
the existence of special drive frequencies
$\omega_D=\omega_m^{\ast}$ for which $f(\gamma_0)=0$. These
correspond to $\gamma_0=\pi \zeta_m/2$ and $\gamma_0=m \pi$
respectively for the cosine and square-pulse protocols, where $m$ is
a positive integer and $\zeta_m$ denotes the position of the $m^{\rm
th}$ zero of $J_0$. This yields
\begin{eqnarray}
\omega_m^{\ast} &=& V_1/(\hbar \zeta_m) \quad {\rm
for} \,\, {\rm cosine}\, \,{\rm  protocol} \nonumber\\
&=& V_1/(2 \hbar m)  \quad {\rm for} \,\, {\rm square-pulse}\, \,
{\rm protocol} \label{freqexp}
\end{eqnarray}
At these frequencies, $H_F^{(1)}$ reduces to the constrained hopping
Hamiltonian studied in Refs.\ \onlinecite{hsf5,hsf6} which is known
to show strong HSF. Such constrained Hamiltonian hosts several
conserved quantities, namely, the Fermion density $N= \sum_j \hat
n_j/L$, $N_d= \sum_j \hat n_j \hat n_{j+1}$, and $N'_d= \sum_j
(-1)^j \hat n_j \hat n_{j+1}$; the corresponding Hilbert space
splits into exponentially large number of fragments including
exponentially large number of frozen states $|\psi_f\rangle$.
Furthermore, in the high-drive amplitude regime, a straightforward
calculation detailed in Ref.\ \onlinecite{supp1} shows that the
higher order terms of $H_F$ are suppressed by at least a factor of
$J/V_1$. Thus in this regime, we expect the driven Fermion chain to
show signatures of HSF in the prethermal regime at $\omega_D=
\omega_m^{\ast}$. For the rest of this work, we present our
numerical results using the square-pulse protocol; the corresponding
results for the cosine drive protocol is presented in Ref.\
\onlinecite{supp1}.

\begin{figure}
\includegraphics[width=\linewidth]{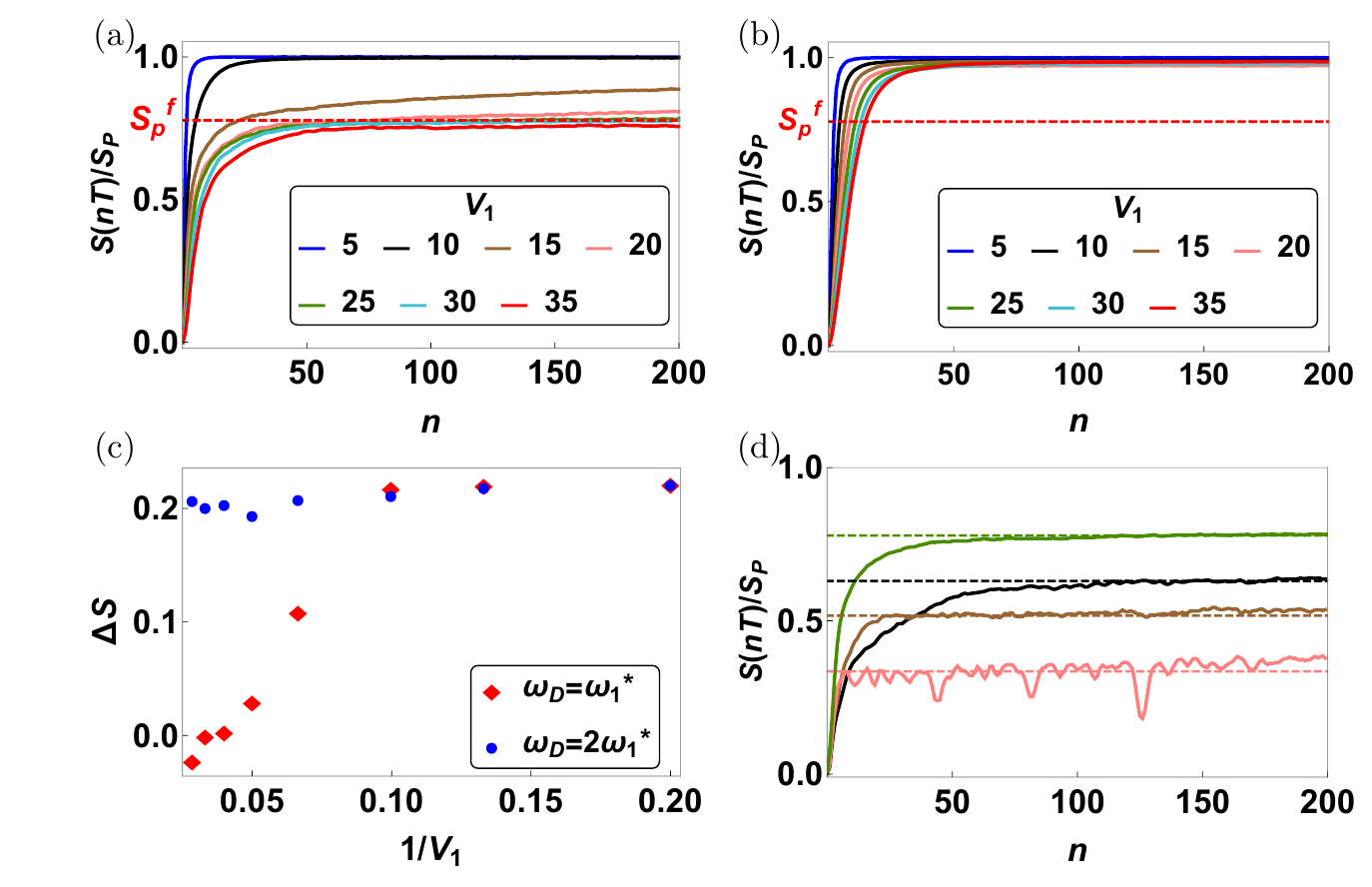}
\caption{(a) Plot of $S(nT)/S_{p}$ as a function of $n$ for
$\omega_1^{\ast}$ and several $V_1$ starting from a random Fock
state. The red dotted line corresponds to $S_p^f/S_p$ of the largest
fragment of $H_F^{(1)}$ with HSD $1008$ to which the initial state
belongs. $S(nT)$ saturates to $S_p^f=0.8 S_p$ for large $V_1$ and $n
\le 200$. (b) Same as (a) but at $2 \omega_1^{\ast}$; here $S$
saturates to $S_p$ for all $V_1$. (c) Plot of $\Delta
S=(S(nT)-S_p^f)/S_{p}$ as a function of $1/V_1$ for $n=200$. $\Delta
S \to 0$ at $\omega_1^{\ast}$ and for large $V_1$. (d) Plot of
$S(nT)/S_{p}$ for $V_1=25$ and at $\omega_1^{\ast}$ corresponding to
several initial Fock states belonging to fragments of $H_F^{(1)}$
with HSD $1008$(green), $288$(black), $144$(brown) and $56$(pink).
The dotted lines correspond to $S_p^f/S_p$ to which $S(nT)/S_p$
saturates for $n\le 200$. All plots indicate clear signature of
prethermal strong HSF at $\omega_1^{\ast}$. For all plots $L=16$,
$V_0= V_2=2$ and all energies are scaled in units of $J$.}
\label{fig1}
\end{figure}

{\it Entanglement entropy}: To find signature of fragmentation in
the driven chain, we first study $S(nT)$ starting from a random Fock
state $|\psi_0\rangle$ as a function of the number of drive cycles
$n$. To this end, we numerically compute $U(T,0)$ using ED; the
details of the procedure is outlined in Ref.\ \onlinecite{supp1}.
This allows us to obtain $|\psi(nT)\rangle = U(nT,0)
|\psi_0\rangle$. One then constructs the density matrix $\rho(nT) =
|\psi(nT)\rangle \langle \psi(nT)|$ for the driven chain. Finally
one carries out a partial trace of $\rho(nT)$ over half the chain,
to obtain the reduced density matrix $\rho_{\rm red} (nT)$
\cite{flscar1}; this leads to $S(nT) = - {\rm Tr} [\rho_{\rm red}
(nT) \ln \rho_{\rm red}(nT)]$.

For an ergodic chain, one expects $S(nT)$ to saturate to the
symmetry resolved Page value $S_{p}$ \cite{page1} corresponding to
the symmetry sector to which the initial state belongs. In contrast,
for a system with HSF, it saturates to the Page value $S_p^f$ of the
fragment of $H_F^{(1)}$ to which the initial state belongs; since
for strong HSF the Hilbert space dimension (HSD) of any fragment is
exponentially smaller than the total HSD, $S_p^f < S_p$.

In Fig.\ \ref{fig1}, plots of $S(nT)$ corresponding to the symmetry
sector $N= \sum_j n_j/L=1/2$ are shown. For these plots we have
chosen $V_0=V_2= 2J$ and used periodic boundary condition. Fig.\
\ref{fig1}(a) shows that for $\omega_D=\omega_1^{\ast}$ and large
$V_1/J$, $S(nT)$ saturates to $S_p^f \simeq 0.8 S_p$ till $n\sim
200$. In contrast, it saturates to $S_p$ for small $V_1/J$. We find
a clear crossover between these two regimes. This behavior is to be
contrasted with its counterpart at $2\omega_1^{\ast}$ (Fig.\
\ref{fig1}(b)), where $S(nT)$ always saturates to $S_p$ within $n
\le 50$. The dependence of $\Delta S=(S(nT)-S_p^f)/S_p$ on $1/V_1$
for both $\omega_1^{\ast}$ (red dots) and $2 \omega_1^{\ast}$ (blue
dots) for $n=200$ is shown in Fig.\ \ref{fig1}(c). The latter always
stays finite indicating proximity to $S_p$ while the former sharply
drops to zero beyond a critical drive amplitude. Finally, we compute
$S(nT)/S_p$ for several random Fock states which belong to different
Hilbert space fragments of $H_F^{(1)}$. We find that for $V_1/J=25$
and at $\omega_1^{\ast}$, $S(nT)$ for these initial states saturate
to their respective $S_p^f$. All these features show a clear
signature of prethermal HSF at $\omega_1^{\ast}$.

\begin{figure}
\includegraphics[width=\linewidth]{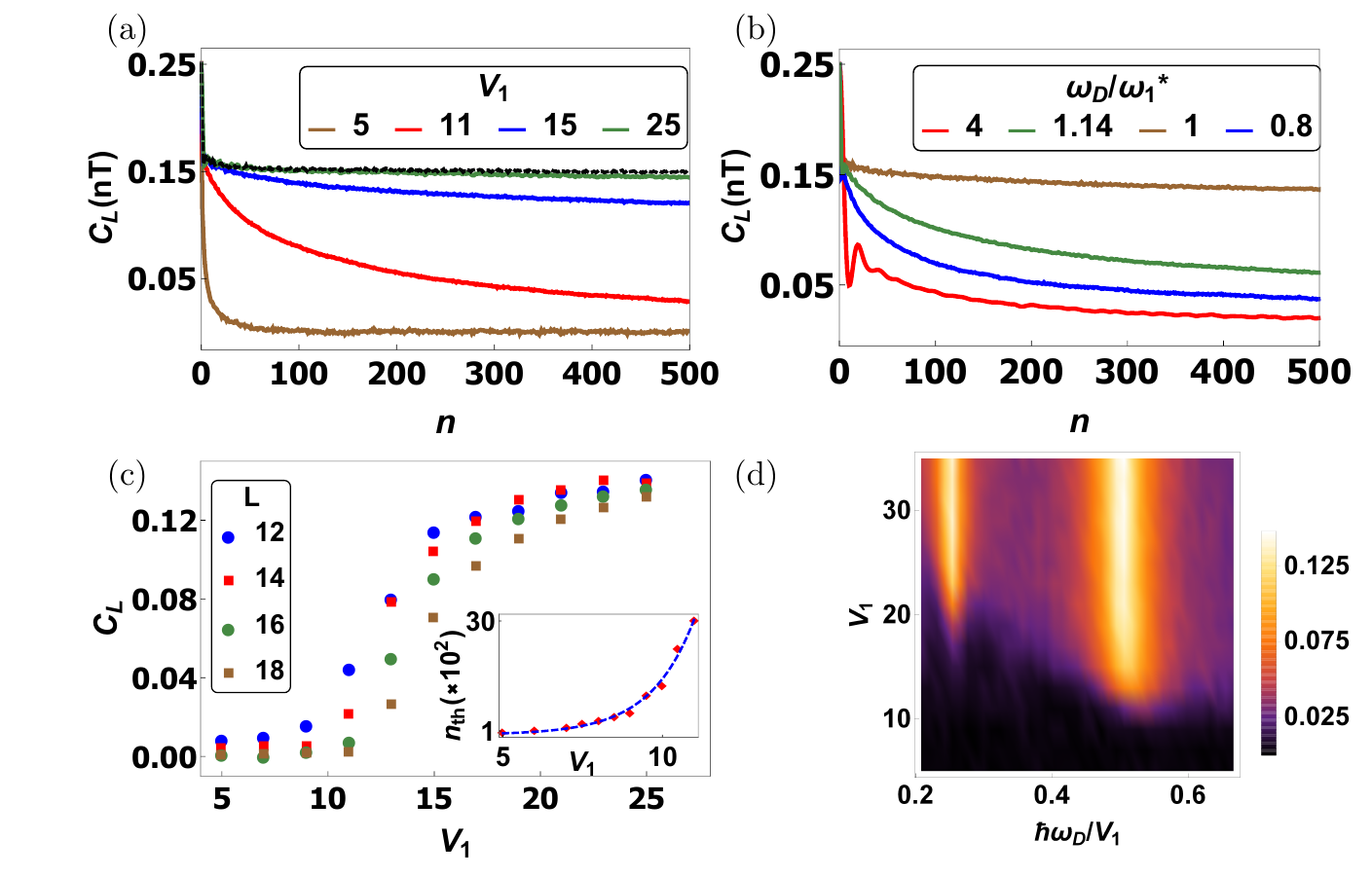}
\caption{(a) $C_L(nT)$ as a function of $n$ at $\omega_1^{\ast}$
computed using exact $H_F$ for several $V_1$ starting from a random
thermal state. For large $V_1$, $C_L(nT)$ saturates to the black
dotted line which corresponds to its value computed using
$H_F^{(1)}$ indicating prethermal HSF. (b) Plot of $C_L(nT)$ as a
function of $n$ for several $\omega_D$ for $V_1=19$ showing ETH
predicted thermalization away from $\omega_1^{\ast}$. (c) Plot of
$C_L(nT)$ for $n=5000$ as a function of $V_1$ at $\omega_1^{\ast}$
for several $L$ showing a clear crossover to a prethermal HSF regime
at large $V_1$. The inset shows the number of cycles, $n_{\rm th}$,
required for $C_L(nT)$ to reach its ETH predicted value at
$\omega_1^{\ast}$ for $L=16$; $n_{\rm th}$ scales exponentially with
$V_1$ showing a long prethermal regime at large $V_1$. d) Phase
diagram obtained from plot of $C_L(nT)$ for $n=5000$ as a function
of $V_1$ and $\hbar \omega_D/V_1$ showing clear signature of finite
value of $C_L$ at large $V_1$ at $\hbar
\omega_{1(2)}^{\ast}=V_1/2(4)$. For all plots $V_0= V_2=2$ and all
energies are scaled in units of $J$. $L=16$ for (a)and (b) and
$L=14$ for (d).} \label{fig2}
\end{figure}

{\it Autocorrelation}: For further signature of HSF at large $V_1$,
we study the autocorrelation function $C_{j=L}(nT) \equiv C_L(nT)$
(Eq.\ \ref{acor}) using ED as a function of $n$. For this purpose,
we use open boundary condition and set $V_0=V_2=2J$. It is
well-known that the presence of HSF leads to a finite long-term
value of the autocorrelator which is bounded from below
\cite{mazur1,hsf2,hsf3,supp1}; for the present system, this bound is
estimated to be $0.125$ \cite{supp1}. In contrast it decays to its
ETH predicted value, $C_{\rm ETH}=0$, in the absence of HSF. The
behavior of $C_j(nT)$ for $j\ne L$ is qualitatively similar
\cite{hsf3,supp1}.

The results for $C_L(nT)$ are shown in Fig.\ \ref{fig2}. Fig.\
\ref{fig2}(a) shows the behavior of $C_L(nT)$ as a function of $n$
at $\omega_1^{\ast}$ for several $V_1/J$. We again find that for
large $V_1/J$, $C_L(nT)$ stays above the lower bound $0.125$ and
close to its value predicted by $H_F^{(1)}$ exhibiting strong HSF
for a large number of drive cycles. The behavior of $C_L(nT)$ for
other drive frequencies are shown in Fig.\ \ref{fig2}(b) for $V_1=25
J$; the plot clearly indicates that deviation from $\omega_1^{\ast}$
leads to rapid, ETH predicted, thermalization of $C_L(nT)$.

The plot of $C_L(nT)$ for $n=5000$ at $\omega_1^{\ast}$ is shown as
a function of $V_1/J$ in Fig.\ \ref{fig2}(c) for several system
sizes ($L$). We find $C_L$ becomes almost independent of $L$ at both
large and small $V_1/J$; in between, the crossover region reduces in
width with increasing $L$. This may indicate a sharp transition in
the thermodynamic limit; however, it is difficult to conclude this
from the present data.

The inset of Fig.\ \ref{fig2}(c), shows a plot of $n_{\rm th}$, the
number of drive cycles required for $C_L$ to reach the ETH predicted
value, as a function of $V_1/J$. The data clearly demonstrates
exponential scaling of $n_{\rm th}$ with $V_1/J$
\cite{scaling1,scaling2,scaling3}. A numerical fit suggests $n_{\rm
th} \sim \exp[0.72 V_1/J]$. This behavior of $n_{\rm th}$ can be
understood as follows. It is expected that for large $\omega_D$, the
extent of the prethermal regime scales exponentially with $\omega_D$
: $n_{\rm th} \sim \exp[c_0 \hbar \omega_D/J]$ where $c_0$ is a
constant \cite{scaling1}. In the present case, $\omega_D=
\omega_1^{\ast}$; so one expects $n_{\rm th} = \exp[c_0 V_1/(2 J)]$
which yields the exponential scaling. The numerical value of $c_0$
is determined to be order unity from our numerics; an analytic
estimate of $c_0$, which necessitates information about breakdown of
convergence of the FPT series, is beyond the scope of the present
work. However, we would like to point out that this exponential
scaling indicates a long and stable prethermal regime where HSF
signatures can be found.

Finally, in Fig.\ \ref{fig2}(d), we show a phase diagram
distinguishing between regimes displaying ETH predicted thermal and
HSF features of $C_L$ as a function of $V_1/J$ and $\hbar
\omega_D/V_1$. The plot represents a clear crossover between the two
regimes at both $\omega_1^{\ast}= V_1/(2\hbar)$ and
$\omega_2^{\ast}=V_1/(4\hbar)$ as $V_1/J$ is increased; we note that
this is consistent with our theoretical expectation based on
$H_F^{(1)}$.

\begin{figure}
\includegraphics[width=\linewidth]{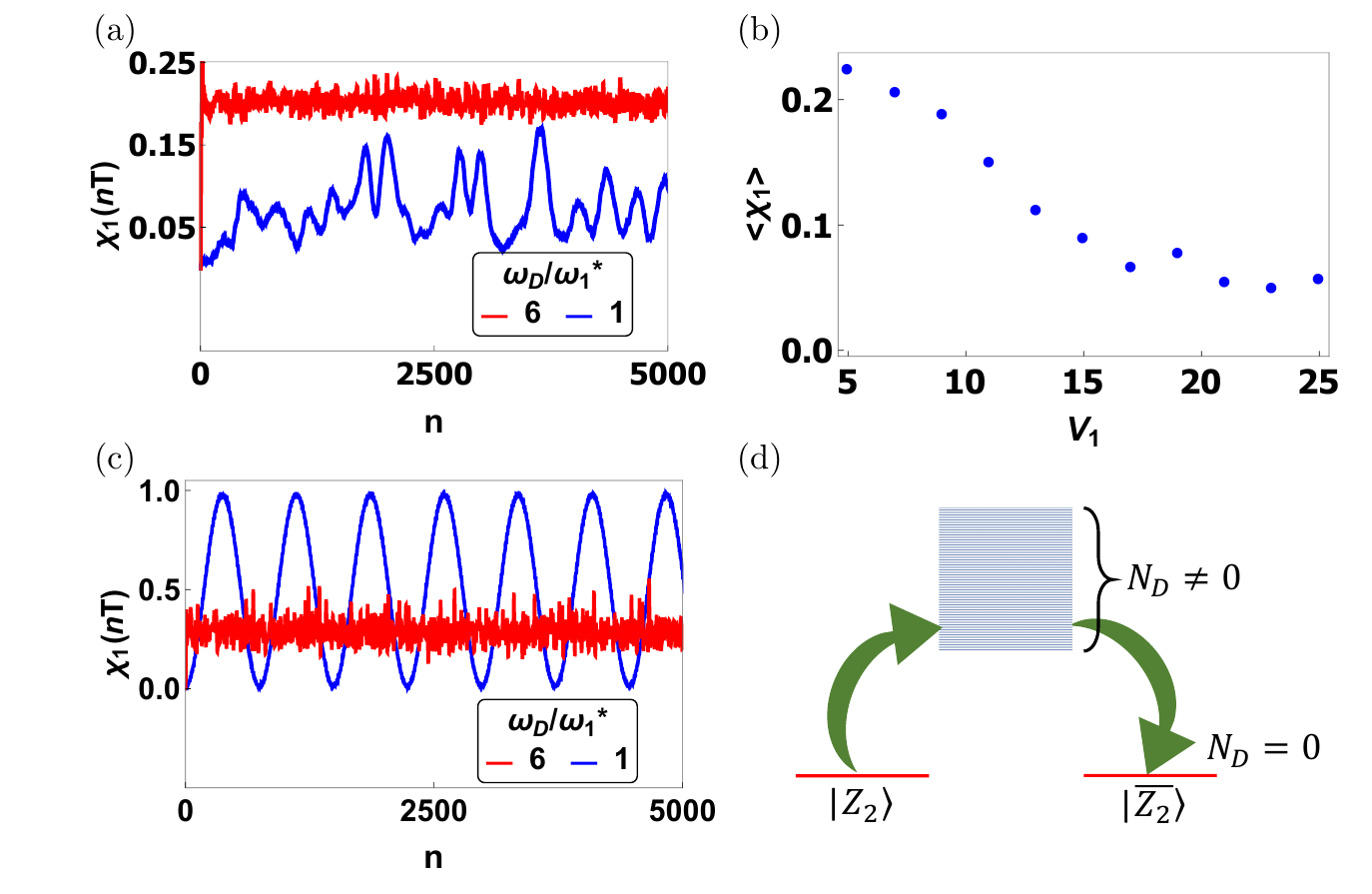}
\caption{(a) Plot of $\chi_1(nT)$ as a function of $n$ for $V_1=19$
at $\omega_1^{\ast}$ (blue curve) and $6 \omega_1^{\ast}$ (red
curve) starting from a random frozen state showing lack of ETH
predicted thermalization at $\omega_1^{\ast}$. (b) Plot of $\langle
\chi_1 \rangle$ as a function of $V_1$ at $\omega_1^{\ast}$;
$\langle \chi_1\rangle$ stays close to its initial value for large
$V_1$ which is consistent with prethermal HSF. (c) Same as in (a)
but for initial $|Z_2\rangle$ state showing slow oscillations at
$\omega_1^{\ast}$. (d) Schematic diagram for the Floquet
quasienergies showing doubly degenerate $|Z_2\rangle$ and $|{\bar
Z}_2\rangle$ with $N_d=0$ and other states with $N_d\ne 0$. The
arrows indicate transition to $|{\bar Z}_2\rangle$ from
$|Z_2\rangle$ using intermediate states with $N_d \ne 0$ leading to
slow oscillations. For all plots $V_0= 10 V_2=2$, $L=14$, and all
energies are scaled in units of $J$.} \label{fig3}
\end{figure}

{\it Dynamics of frozen states}: The frozen states correspond to
Fock states which are eigenstates of $H_F^{(1)}$; however, they have
non-trivial evolution under $H_F$ due to higher order terms in FPT.
The nature of this evolution, for a random frozen state with no
particular symmetries, is expected to obey ETH away from
$\omega_m^{\ast}$; in contrast, at $\omega_m^{\ast}$, they are
expected to stay close to their initial values at large drive
amplitudes for a wide range of $n$.

To capture this behavior we plot $\chi_{j=1}(nT) \equiv \chi_1(nT)$
(Eq.\ \ref{etcorr}) as a function of $n$ in Fig.\ \ref{fig3}(a) for
$\omega_1^{\ast}$ and $6\omega_1^{\ast}$ at $V_1=19 J$. These
computations are done for chains with periodic boundary conditions
and $V_0=10 V_2=2J$. The plot clearly shows that $\chi_1(nT)$
reaches its ETH predicted value for $6 \omega_1^{\ast}$; in contrast
it never reaches the ETH predicted value at $\omega_1^{\ast}$. The
plot of $\langle \chi_1 \rangle$ ($\chi_1(nT)$ averaged over $5000$
drive cycles starting from $n=5000$) as a function of $V_1/J$ at
$\omega_1^{\ast}$ is shown in Fig.\ \ref{fig3}(b). We find that
$\langle \chi_1\rangle$ stays close to its initial value at large
$V_1/J$ which is consistent with prethermal strong HSF. We have
checked that this behavior is similar for all $\langle
\chi_j\rangle$.

Next we study the dynamics of frozen state when $|\psi_f\rangle=
|Z_2\rangle$. As shown in Fig.~\ref{fig3}(c), when the system is
driven at $\omega_1^{\ast}$ (blue curve), the equal-time correlation
function $\chi_1(nT)$ shows slow oscillations with a time period
that is orders of magnitude longer than the bare time scales implied
in the Hamiltonian of Eq.~\eqref{fermham}. On the other hand, the
oscillations are absent for $6 \omega_1^{\ast}$ (red curve in
Fig.~\ref{fig3}(c)). In fact, the occurrence of the slow
oscillations require two conditions to be satisfied. First, the
system has to be fragmented in a prethermal sense so that, starting
from the state $|Z_2\rangle$, over long time scales the system stays
effectively confined in the $N_D=0$ sector with no nearest neighbor
occupations. This sector comprises the states $|Z_2\rangle$ and
$|\bar{Z}_2\rangle \equiv |1, 0, 1, 0, \cdots \rangle$. Second, the
energy scales $J/V_0\lesssim 1$ and $V_2/V_0 \le 1/2$, so that
$N_D=0$ is the lowest energy manifold in the Fock space, while $N_D
\neq 0$ are high energy states as shown in Fig.~\ref{fig3}(d). In
such a situation the higher order terms in FPT provide tunneling
paths for the system to oscillate between $|Z_2\rangle$ and
$|\bar{Z}_2 \rangle$. Since $\hat{n}_1 \hat{n}_3 | Z_2 \rangle =0$,
while $\hat{n}_1 \hat{n}_3 | \bar{Z}_2 \rangle =1$, the equal-time
correlation function $\chi_1(nT)$ oscillates between zero and nearly
one (the deviation from one is due to finite mixing with the states
in the $N_D \neq 0$ sectors). In other words, the oscillations are
manifestation of the tunneling processes that restore $Z_2$ symmetry
in a finite system such that the approximate eigenstates of $H_F$
are the bonding and antibonding states $|\psi_{B, A} \rangle \equiv
| Z_2 \rangle \pm | \bar{Z}_2 \rangle$. Therefore, the oscillation
frequency is proportional to the energy split $2\hbar \alpha_d$
between the bonding and antibonding states; this is expected to be a
small energy scale as it arises from higher order terms in $H_F$
(which we verified using ED). Thus, if $H_F |\psi_{B, A} \rangle
\approx \hbar (\alpha_s \pm \alpha_d) |\psi_{B, A} \rangle$ for a
constant $\alpha_s$, one can show that $\chi_1(nT) \approx \sin^2
(\alpha_d nT)$ which explains the slow oscillations. In passing we
note that similar oscillations also appear in the fidelity function
$F(nT) \equiv | \langle Z_2(0) |Z_2(nT) \rangle |^2$
[\onlinecite{supp1}].

{\it Discussion}: In conclusion, we have shown that a driven Fermion
chain, in the high drive amplitude regime, exhibits signature of
strong HSF in the prethermal regime at special drive frequencies.
This constitutes the first realization of strong HSF in
out-of-equilibrium systems. We have provided analytic expressions
for these special frequencies and have supported this claim by exact
numerical computation of entanglement entropy, autocorrelation and
equal-time correlation function of the Fermions. We have also
studied the dynamics of these Fermions starting from frozen states
and have identified a novel oscillatory dynamics for the
$|Z_2\rangle$ initial state; such oscillations arise due to both HSF
and $Z_2$ symmetry breaking. We expect our results to be of
relevance for ultracold atom platforms where such Fermion chains may
be experimentally realized \cite{exprev,hsf7}.

{\it Acknowledgement}: SG acknowledges CSIR, India for support
through project 09/080(1133)/2019-EMR-I. IP thanks Edouard Boulat
for discussions. KS thanks DST, India for support through SERB
project JCB/2021/000030 and Arnab Sen for discussions.

\appendix

\section{Floquet Perturbation Theory}
\label{fpt}

In this section, we provide the details of the Floquet perturbation
theory computations outlined in the text for both the cosine and the
square-pulse protocols. To this end, we first note that in the large
drive amplitude limit, where $V_1 \gg V_0, J, V_2$, one can write
the Hamiltonian $H(t)= H_0(t) + H_1$, where
\begin{eqnarray}
H_0(t) &=& V(t) \sum_j \hat n_j \hat n_{j+1}  \nonumber\\
H_1 &=& -J \sum_{j} (c_j^{\dagger} c_{j+1} + {\rm h.c}) + V_0
\sum_{j} \hat n_j \hat n_{j+1} \nonumber\\
&& + V_2 \sum_j \hat n_j \hat n_{j+2}  \label{ham1}
\end{eqnarray}
where $V(t)= V_1 f(t)$ and $f(t)$ is a periodic function with time
period $T$. The precise form of $f(t)$ depends on the protocol as
outlined in the main text; for the cosine-drive protocol, $f(t) =
\cos \omega_D t$, where $\omega_D=2 \pi/T$ is the drive frequency
while for the square-pulse protocol $f(t)=-1(1)$ for $t\le(>) T/2$.

Starting from Eq.\ \ref{ham1}, we first construct the evolution
operator
\begin{eqnarray}
U_0(t,0) &=& {\mathcal T} \left[ e^{-i \int_0^t dt' H_0(t')/\hbar}
\right]  \label{evol0}
\end{eqnarray}
Using Eq.\ \ref{ham1}, this can be evaluated to obtain
\begin{eqnarray}
U_0^c(t,0) &=& e^{ -i V_1 \sin\omega_D t \sum_j \hat n_j \hat
n_{j+1}/(\hbar \omega_D)} \label{cosu0}
\end{eqnarray}
for the cosine drive protocol and
\begin{eqnarray}
U_0^s(t,0) &=& e^{i V_1 t \sum_j \hat n_j \hat n_{j+1}/\hbar}, \quad
t \le T/2  \nonumber\\
&=& e^{i V_1 (T-t) \sum_j \hat n_j \hat n_{j+1}/\hbar}, \quad t >
T/2 \label{squ0}
\end{eqnarray}
for the square-pulse protocol. Note that for both protocols
$U_0^c(T,0)= U_0^s(T,0)= I$, where $I$ denotes the identity matrix;
this leads to $H_F^{(0)}=0$ for both protocols.

To obtain the first order Floquet Hamiltonian, we now treat $H_1$
perturbatively following Refs.\ \onlinecite{rev8,flscar2}. First, we
consider the continuous protocol for which the first order
correction to $U_0^c$ is given by
\begin{eqnarray}
U_1^c(T,0) &=& \frac{-i}{\hbar} \int_0^T dt (U_0^{c}(t,0))^{\dagger}
H_1 U_0^c(t,0) \label{foc}
\end{eqnarray}
To evaluate $U_1^c$, we note that the terms in $H_1$ involving $V_0$
and $V_2$ (Eq.\ \ref{ham1}) commute with $H_0$ and hence with
$U_0^c$ while the hopping term does not. To understand the effect of
the hopping term, we consider the action of $U_0^c$ on an arbitrary
Fock state $|m\rangle$. Since $H_0(t)$ is diagonal in the Fock
basis, one can write
\begin{eqnarray}
U_0^c(t,0) |m\rangle &=& e^{-i E_m \sin \omega_D t/(\hbar \omega_D)}
|m\rangle \equiv e^{-i \theta_m} |m\rangle, \nonumber\\
V_1 \sum_j \hat n_j \hat n_{j+1} |m\rangle &=& E_m |m\rangle.
\label{uomat}
\end{eqnarray}
Now let us consider the action of the hopping term in $H_1$ on the
state $|m\rangle$: $-J c_j^{\dagger} c_{j+1} |m\rangle = |n\rangle$.
We note that $|m\rangle$ and $|n\rangle$ have different occupation
numbers only on the $j^{\rm th}$ and $(j+1)^{\rm th}$ sites. It is
then easy to see that
\begin{eqnarray}
E_m &=& E_n,  \quad  {\rm if}  \, \, A_j |m\rangle =0  \nonumber\\
&=& E_n \pm V_1, \quad {\rm if} \, \, A_j |m\rangle = \pm |m\rangle
\nonumber\\
A_j &=& (\hat n_{j+2} - \hat n_{j-1})  \label{hopp1}
\end{eqnarray}
The result of Eq.\ \ref{hopp1} can be understood in the following
manner. We find that the phase $\theta_m $ of any Fock state can
change due to local hopping if this process changes the number of
nearest neighbor pairs; such a change can only occur if $A_j$ is
non-zero. Using this, one can carry out the computation of $U_1^c$
following standard methods outlined in Ref.\
\onlinecite{rev8,flscar2}. This yields
\begin{eqnarray}
U_1^c(T,0) &=& \frac{-i T}{\hbar} \sum_j \hat n_j (V_0 \hat n_{j+1}
+
V_2 \hat n_{j+2}) \label{u1eq1} \\
&&+\frac{i T J}{\hbar}\sum_j \left[J_0\left( \frac{2 \gamma_0
A_j}{\pi} \right) c_j^{\dagger} c_{j+1} + \rm {h.c} \right]
\nonumber
\end{eqnarray}
where $\gamma_0= V_1 T/(4\hbar)$, $J_0$ denotes the zeroth order
Bessel function and we have used the identity $\exp[i \alpha \sin
x]= \sum_m J_m(\alpha) \exp[i m x]$. Next using the properties
$J_0(x)= J_0(-x)$ and $J_0(0)=1$ and noting that $A_j$ can only have
values $0, \pm 1$ for any $j$, we write
\begin{eqnarray}
J_0\left( \frac{2\gamma_0 A_j}{\pi}\right) &=& J_0\left(
\frac{2\gamma_0}{\pi}\right) A_j^2 + (1-A_j^2) \label{besid}
\end{eqnarray}
This allows us to write
\begin{eqnarray}
&&U_1^c(T,0) = \frac{-i T}{\hbar} \sum_j \hat n_j (V_0 \hat n_{j+1}
+
V_2 \hat n_{j+2}) \label{u1eq2}\\
&&+\frac{i T J}{\hbar}\sum_j \left[\left(J_0\left(
\frac{2\gamma_0}{\pi}\right)A_j^2  +(1-A_j^2)\right) c_j^{\dagger}
c_{j+1} + \rm {h.c} \right]\nonumber
\end{eqnarray}
This leads to $ H_F^{(1,c)}= i \hbar U_1(T,0)/T$ which is given by
\begin{eqnarray}
&& H_F^{(1,c)}= \sum_j \hat n_j (V_0 \hat n_{j+1} +
V_2 \hat n_{j+2})\label{heff1c} \\
&&-J\sum_j \left[\left(J_0\left( \frac{2\gamma_0}{\pi}\right)A_j^2
+(1-A_j^2)\right) c_j^{\dagger} c_{j+1} + \rm {h.c} \right]\nonumber
\end{eqnarray}
which is the result used in the main text.

Next, we consider the second order term. To second order, the
evolution operator is given by
\begin{eqnarray}
U_2^c(T,0) &=& \left(\frac{-i}{\hbar}\right)^2 \int_0^T dt_1 U_0^{c
\dagger}(t_1,0) H_1 U_0^c(t_1,0) \nonumber\\
&& \times \int_0^{t_1} dt_2 U_0^{c \dagger}(t_2,0) H_1 U_0^c(t_2,0).
\label{u2eqa}
\end{eqnarray}
The second order contribution to the Floquet Hamiltonian is given,
in terms of $U_2^c$ by
\begin{eqnarray}
H_F^{(2,c)}(T,0) &=& \frac{i\hbar}{T} (U_2^c(T,0) -
[U_1^c(T,0)]^2/2) \label{heff2a}
\end{eqnarray}
From Eqs.\ \ref{heff2a} and \ref{u2eqa}, it is easy to see that
there is no contribution to $H_F^{(2,c)}$ from terms in $H_1$ which
commutes with $U_0^c$; their contributions to $U_2^c$ and
$(U_1^c)^2/2$ cancel each other. Thus the only contribution to
$H_F^{(2)}$ comes from the hopping term. A straightforward, albeit
somewhat lengthy, calculation yields
\begin{eqnarray}
H_F^{(2,c)}(T,0) &=& -\frac{2J^2}{\hbar \omega_D}
C_0\left(\frac{2\gamma_0}{\pi}\right) \sum_{j_1,j_2}
\left[O^{(1)}_{j_1},O_{j_2}^{(2)}\right] \nonumber\\
C_0(\alpha) &=& \sum_{m=0}^{\infty} J_{2m+1}(\alpha)/(2m+1)
\nonumber\\
O_{j_1}^{(1)} &=& \Big[ \left (J_0\left(\frac{2\gamma_0}{\pi}\right)
A_{j_1}^2 + (1-A_{j_1}^2)\right)
\nonumber\\
&& \times c_{j_1}^{\dagger}
c_{j_1+1} +{\rm h.c.} \Big] \nonumber\\
O_{j_2}^{(2)} &=& A_{j_2} c_{j_2+1}^{\dagger} c_{j_2} -{\rm h.c.}
\end{eqnarray}
where we have used the fact that $J_{2m+1}(-x)=- J_{2m+1}(x)$ and
$J_{2m+1}(0)=0$. The commutator structure of $H_F^{(2,c)}$ ensures
that it is local. Furthermore, $||H_F^{(2,c)}|| < ||H_F^{(1,c)}||$
for $J/(\hbar \omega_D) <1$; this ensures that one can use this
perturbative scheme in the intermediate frequency regime, where
$\hbar \omega_D < V_1$. However, we note that the above-mentioned
criteria does not ensure that the perturbation series shall converge
for $\hbar \omega_D \sim J$; estimating the radius of convergence of
such an expansion is beyond the scope of the present work.

Next we consider the square pulse protocol. For this the first order
Floquet Hamiltonian can be written as
\begin{widetext}
\begin{eqnarray}
U_1^s(T,0) &=& \frac{-i}{\hbar} \left(\int_0^{T/2} dt e^{-i V_1 t
\sum_j \hat n_j \hat n_{j+1}/\hbar} H_1 e^{i V_1 t \sum_j \hat n_j
\hat n_{j+1}/\hbar} + \int_{T/2}^T e^{-i V_1 (T-t) \sum_j \hat n_j
\hat n_{j+1}/\hbar} H_1 e^{i V_1 (T-t) \sum_j \hat n_j \hat
n_{j+1}/\hbar} \right) \nonumber\\ \label{u1sq}
\end{eqnarray}
\end{widetext}
Eq.\ \ref{u1sq} can be evaluated exactly as done for the continuous
protocol. We define the phase $\theta'_m$ via the relation
$U_0^s(t,0)|m\rangle = \exp[i \theta'_m] |m\rangle$ and then note
that the action of a local hopping term can change this phase only
if it changes the number of nearest neighbor pairs. This condition
can again be implemented using the operator $A_j$. This leads to
\begin{eqnarray}
&& U_1^s(T,0) = \frac{-i T}{\hbar} \sum_j \hat n_j (V_0 \hat n_{j+1}
+ V_2 \hat n_{j+2}) \label{u1seq2}\\
&&+\frac{i T J}{\hbar} \sum_j \frac{ \sin \gamma_0 A_j}{\gamma_0
A_j} e^{i \gamma_0 A_j} c_{j}^{\dagger} c_{j+1} + \rm {h.c}
\nonumber
\end{eqnarray}
Noting that $A_j=0, \pm 1$, one can then write
\begin{eqnarray}
\frac{ \sin \gamma_0 A_j}{\gamma_0 A_j} e^{i \gamma_0 A_j} &=&
\frac{ \sin \gamma_0}{\gamma_0} A_j^2 e^{i \gamma_0 A_j} + (1-A_j^2)
\label{rel2}
\end{eqnarray}
Using Eq.\ \ref{rel2} and the relation $H_F^{(1,s)} = i \hbar
U_1^s(T,0)/T$, one finally gets
\begin{widetext}
\begin{eqnarray}
H_F^{(1,s)}(T,0) &=& \sum_j \hat n_j (V_0 \hat n_{j+1} + V_2 \hat
n_{j+2}) -J \sum_j \left(\left[\frac{ \sin \gamma_0}{\gamma_0} e^{i
\gamma_0 A_j} A_j^2 +(1-A_j^2)\right] c_{j}^{\dagger} c_{j+1} + \rm
{h.c} \right) \label{hf1s}
\end{eqnarray}
\end{widetext}
which yields the result of the main text.

The second order contribution to $H_F^{(2)}$ for the square-pulse
protocol vanishes. This can be seen as follows. The second order
evolution operator for the square pulse protocol is given
analogously by
\begin{eqnarray}
U_2^s(T,0) &=& \left(\frac{-i}{\hbar}\right)^2 \int_0^T dt_1 U_0^{s
\dagger}(t_1,0) H_1 U_0^s(t_1,0) \nonumber\\
&& \times \int_0^{t_1} dt_2 U_0^{s \dagger}(t_2,0) H_1 U_0^s(t_2,0).
\label{u2sq}
\end{eqnarray}
and the corresponding second order Floquet Hamiltonian reads
\begin{eqnarray}
H_F^{(2,s)}(T,0) &=& \frac{i\hbar}{T} (U_2^s(T,0) -
[U_1^s(T,0)]^2/2) \label{heff2sq}
\end{eqnarray}

Using the form of the zeroth order evolution operator from Eq.\
\ref{squ0}, it can be easily shown that the integral in Eq.\
\ref{u2sq} can be split into three non-zero integrals of the general
form
\begin{widetext}
\begin{eqnarray}
&&I_1=\int_0^Tdt_1\int_0^{t_1}dt_2 \theta(T/2-t_1)\theta(T/2-t_2)f(t_1) g(t_2)\nonumber\\
&&I_2=\int_0^Tdt_1\int_0^{t_1}dt_2 \theta(t_1-T/2)\theta(t_2-T/2)f(T-t_1) g(T-t_2)\\
&&I_3=\int_0^Tdt_1\int_0^{t_1}dt_2
\theta(t_1-T/2)\theta(T/2-t_2)f(T-t_1) g(t_2)\nonumber
\label{intusq2}
\end{eqnarray}
\end{widetext}
with $\theta(t)$ being the Heaviside step function and $f(t)$,
$g(t)$ being well-behaved functions of $t$, whose exact form depends
on the details of $H_1$ and $U_0^s(t,0)$.

By shifting the limits of these integrals appropriately, one can
show that
\begin{equation}
    I_1+I_2+I_3=2\int_0^{T/2}d\tau_1 f(\tau_1)\int_0^{T/2}d\tau_2 g(\tau_2),
\end{equation}
thus implying that the second order evolution operator
\begin{eqnarray}
U_2^s(T,0) &=& 2\left(\frac{-i}{\hbar}\right)^2 \int_0^{T/2} dt_1
U_0^{s
\dagger}(t_1,0) H_1 U_0^s(t_1,0) \nonumber\\
&& \times \int_0^{T/2} dt_2 U_0^{s \dagger}(t_2,0) H_1 U_0^s(t_2,0).
\label{u2sqint}
\end{eqnarray}
Using the same logic, the first order evolution operator in Eq.\
\ref{u1sq} has the form $U_1^s(T,0) = 2\left(\frac{-i}{\hbar}\right)
\int_0^{T/2} dt_1 U_0^{s \dagger}(t_1,0) H_1 U_0^s(t_1,0)$, so that
\begin{equation}
    U_2^s(T,0)=\frac{[U_1^s(T,0)]^2}{2}
\end{equation}
which implies from Eq.\ \ref{heff2sq} that $H_F^{(2,s)}(T,0)=0$.
Thus the leading order correction to $H_F^{(1)}$ for the
square-pulse protocol comes from the third-order terms which are
$\sim J^3/V_1^2$; we do not aim to compute them in this work.

\section{Half-Chain Entanglement Entropy Computation and Calculation of Page Value}
\label{hcee}

In this section, we outline the procedure for the numerical
computation of the half-chain entanglement entropy and analytical
calculation of Page value used in Fig.\ \ref{fig1} of the main text.
This procedure is applicable for any arbitrary pure state drawn from
the Hilbert space.

We divide the chain into two halves, $A$ and $B$, such that the full
Hilbert space of the system can be represented as
$\mathcal{H}=\mathcal{H}_A \otimes \mathcal{H}_B$. In the
computational basis, any arbitrary state of the whole chain can be
written as
\begin{equation}
    |\psi\rangle = \sum_{i,j} c_{i,j} |i_A\rangle \otimes |j_B\rangle
    \label{hceestate}
\end{equation}
where $c_{i,j}$, which are elements of a matrix $C$, are, in
general, complex numbers and $\{|i_A\rangle\}$ and $\{|j_B\rangle\}$
span $\mathcal{H}_A$ and $\mathcal{H}_B$ respectively.

Next, using singular value decomposition (also known as Schmidt
decomposition in this particular case), we obtain the matrix $\tilde C$ such that $C=P\Tilde{C}Q^\dagger$.
Here $P$ and $Q$ are unitary matrices and $\Tilde{C}$ is a diagonal
matrix with real, non-negative entries $\tilde c_{kk}$. Substituting
this in Eq.\ \ref{hceestate} we find
\begin{eqnarray}
    |\psi\rangle &=& \sum_{i,j,k} p_{ik} \Tilde{c}_{kk} q^*_{jk} |i_A\rangle \otimes |j_B\rangle\nonumber\\
    &=& \sum_k \Tilde{c}_{kk} |k_A\rangle \otimes |k_B\rangle
    \label{hceestate1}
\end{eqnarray}
where $|k_A\rangle=\sum_i p_{ik} |i_A\rangle$ and $|k_B\rangle
=\sum_j q^*_{jk} |j_B\rangle$ are the transformed basis vectors with
unitary matrices $P^T$ and $Q^\dagger$ respectively.

The density matrix of the full system in this basis reads
\begin{equation}
    \rho = |\psi\rangle\langle\psi|=\sum_{k,k'} \Tilde{c}_{kk} \Tilde{c}_{k'k'} (|k_A\rangle \otimes |k_B\rangle)(\langle k'_B| \otimes \langle k'_A|).
    \label{rhofull}
\end{equation}
In order to obtain the reduced density matrix corresponding to one
of the halves, we trace over the degrees of freedom of the other
half. Without loss of generality, we choose to trace out the degrees
of freedom in $B$. This yields
\begin{eqnarray}
    \rho_{A}&=&\text{Tr}_{B} \rho \nonumber\\
    &=&\sum_{k} \Tilde{c}^2_{kk} |k_{A}\rangle\langle k_{A}|
    \label{rhored}
\end{eqnarray}
The half-chain entanglement entropy $S_{A}$ is then given by
$S_{A}=-\text{Tr}_A \rho_{A} \ln \rho_{A}$. Since the reduced
density matrix, $\rho_{A}$ is diagonal in this basis, its
eigenvalues can be easily read off as $\Tilde{c}_{kk}^2$, which
implies that
\begin{equation}
    S_{A}=-\sum_k \Tilde{c}_{kk}^2 \ln \Tilde{c}_{kk}^2
    \label{ee}
\end{equation}
The advantage of the Schmidt decomposition is that one can directly
use $\Tilde{c}_{kk}$ (which are singular values of the coefficient
matrix $C$) in order to evaluate the half-chain entanglement entropy
numerically, without the need to compute the reduced density matrix
explicitly and diagonalizing it.

To calculate the Page value of the entanglement entropy in the
half-filled symmetry sector, we adopt the procedure outlined in
Ref.\ \onlinecite{page1}. We randomly choose a Gaussian orthogonal
ensemble (GOE)-like state in this sector of the form
\begin{equation}
    |\psi\rangle = \sum_{N_A=0}^{L/2} \sum_{b=1}^{d_B(N_B)}
    \sum_{a=1}^{d_A(N_A)} \frac{z_{a,b}}{\sqrt{\mathcal{D}_{L/2}}} |a;N_A\rangle |b;N_B\rangle
    \label{goestate}
\end{equation}
where $z_{a,b}$ are random real numbers chosen from a normal
distribution with mean $0$ and unit variance, $\mathcal{D}_{L/2}$ is
the Hilbert space dimension (HSD) of the half-filled sector, and
$d_A(N_A)$ and $d_B(N_B)$ are HSD's of subsystems $A$ and $B$ with
occupancies $N_A$ and $N_B = \frac{L}{2} - N_A$ respectively. The
dimensions can be easily calculated using combinatorics as
$\mathcal{D}_{L/2}= {^L}C_{L/2}$, $d_A(N_A)= ^{L/2}C_{N_A}$ and
$d_B(N_B)= ^{L/2}C_{N_B}$.

The reduced density matrix for the subsystem $A$ can be calculated
as
\begin{eqnarray}
\rho_A &=& \frac{1}{\mathcal{D}_{L/2}}\sum_{N_A=0}^{L/2}
\sum_{a,a'=1}^{d_A(N_A)} \mu_{a a'} |a,N_A\rangle \langle a',N_A| \nonumber\\
\mu_{a a'} &=& \sum_{b=1}^{d_B(N_B)} z_{a,b} z_{a',b}
\label{goerhored}
\end{eqnarray}
Averaging over the Gaussian distributed random variables and using
\begin{eqnarray}
\overline {\mu_{a,a'}} &=& \overline{\sum_{b=1}^{d_B(N_B)} z_{a,b}
z_{a',b}}= d_B(N_B)\delta_{aa'}, \label{avrel}
\end{eqnarray}
we get for the average reduced density matrix
\begin{equation}
\Bar{\rho}_A=\sum_{N_A=0}^{L/2} \sum_{a=1}^{d_A(N_A)}
\frac{d_B(N_B)}{\mathcal{D}_{L/2}} |a,N_A\rangle\langle a,N_A|
    \label{goerhoredavg}
\end{equation}
This implies that the mean value of the half-chain entanglement
entropy (averaged over GOE states) is $\Bar{S}_{L/2}=-
\overline{\text{Tr}_A \rho_A \ln \rho_A}$. To estimate the maximal
value of $\Bar{S}_{L/2}$, we use the concavity property of the
function $f(x)= x\ln x$, which implies
\begin{eqnarray}
    \Bar{S}_{L/2}\leq -\text{Tr}_A\Bar{\rho}_A \ln \Bar{\rho}_A = S_{L/2}^{\text{max}}
    \label{smax}
\end{eqnarray}
Hence, the maximal value of the half-chain entanglement entropy for
thermal GOE states is
\begin{equation}
    S_{L/2}^{\text{max}}=-\sum_{N_A=0}^{L/2} \frac{d_A(N_A)d_B(N_B)}{\mathcal{D}_{L/2}}\ln \frac{d_B(N_B)}{\mathcal{D}_{L/2}}
    \label{smax1}
\end{equation}
The leading correction to Eq.\ \ref{smax1} was found to be
-$\frac{1}{2}$ in Ref.\ \onlinecite{page1} for the half-filled case.
This is the result which we use to calculate the Page value, $S_p$
of the entire half-filled sector in Fig.\ \ref{fig1} of the main
text. To calculate the Page value of the largest fragment, $S_p^f$
we execute the same procedure of averaging over random canonical
states from that fragment. In this case, we carry out numerical
computations to obtain $S_p^f$.

\begin{figure}
\includegraphics[width=1.02\linewidth]{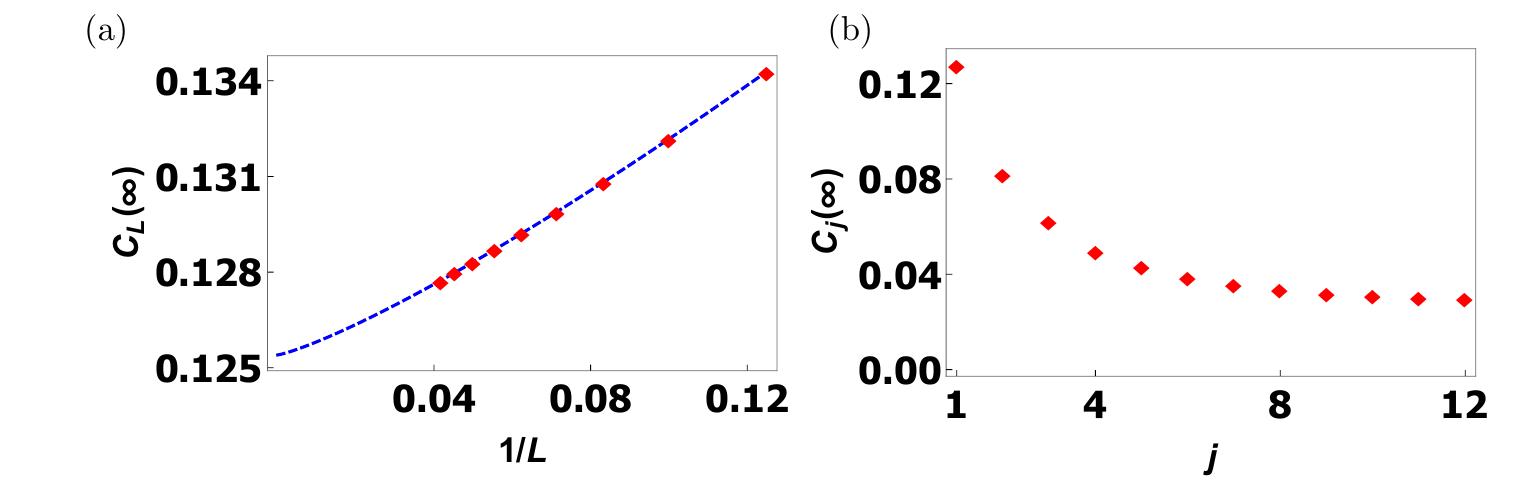}
\caption{(a) Scaling of the lower bound of the fermion density
autocorrelator $C_L(\infty)$, obtained using $H_F^{(1)}$ with system
size, $L$. The blue dashed line corresponds to a least square fit
which predicts a lower bound close to $0.125$ in the thermodynamic
limit. (b) Variation of $C_j(\infty)$ with distance from one of the
edges (labelled as $j=1$) in a chain of length $L=24$. The plot
shows that the bound decreases as one moves away from the edges.}
\label{figs1}
\end{figure}

\section{Lower Bound on the unequal time fermion density autocorrelation function}
\label{mazurbound}

In this section, we establish that in the fragmented regime, the
fermion density autocorrelation function, $C_L(nT)$ computed in the
main text is bounded from below by a value greater than zero.  This
clearly implies that the system evades thermalization (in the
prethermal sense) as the ETH predicted value of the autocorrelator
is zero.

To estimate this lower bound, we consider the fragmented first order
Floquet Hamiltonian, $H_F^{(1)}$, with open boundary condition and
use Mazur's inequality \cite{mazur1}. To this end, we define
${\bar C}_j$ as
\begin{eqnarray}
\bar C_j &=& \lim_{n \to \infty} \frac{1}{n} \sum_n \langle (\hat
n_j(nT)-1/2) (\hat n_j(0)-1/2) \rangle \label{barceq}
\end{eqnarray}
where the expectation on the RHS is taken with respect to an
infinite temperature thermal state. In the presence of HSF, $\bar
C_j$ satisfies
\begin{eqnarray}
\bar C_j\geq \sum_i\frac{[\text{Tr} ((\hat n_j -1/2)
P_i)]^2}{\mathcal{D}_{L/2}\mathcal{D}_i} \equiv C_j(\infty)
\label{mazur}
\end{eqnarray}
where $P_i$ is the projection operator to the $i^\text{th}$ fragment
and $\mathcal{D}_i$ is its dimension, and we shall first set the
coordinate $j=L$ to estimate $C_L(\infty)$. We note that the
presence of higher order terms in $H_F$ at the special drive
frequencies $\omega_m^{\ast}$ will result in violation of this bound
beyond the prethermal regime when $n>n_{\rm th}$; however, as noted
in the main text, in the large drive amplitude regime where the
prethermal regime is long, ${\bar C}_L(nT)$ is expected to remain
above $C_L(\infty)$ for a large range of $n < n_{\rm th} \sim {\rm
O}(e^{V_1/J})$.

To estimate $C_L(\infty)$ using Eq.\ \ref{mazur}, we construct the
matrix for $H_F^{(1)}$ in the computational basis and identify
blocks of connected components. We determine the sizes of these
blocks to obtain ${\mathcal D}_i$s and use these to evaluate
$C_L(\infty)$ for a specific $L$. Using these values of
$C_L(\infty)$ for different $L$, we carry out a $1/L$ extrapolation
as shown in Fig.\ \ref{figs1}(a). We find that $C_L(\infty)$ attains
a finite non-zero value, close to $0.125$ in the thermodynamic
limit, thus affirming non-thermal behavior. This is consistent with
the results obtained in Fig.\ \ref{fig2} of the main text where we
find $C_L(nT) \simeq 0.15$ in the prethermal regime at
$\omega_D=\omega_1^{\ast}$. We also study the variation of this
lower bound with distance from one of the edges of the chain in
Fig.\ \ref{figs1}(b), where we plot $C_j(\infty)$ for $L=24$ as a
function of $j$. We find that the bound decreases as one moves into
the bulk. This behavior is consistent with that found in Refs.\
\onlinecite{hsf2,hsf3} and is the reason why $C_L(\infty)$ is
studied in details in the main text.

\section{Fidelity}
\label{fid}
\begin{figure}
\includegraphics[width=1.0\linewidth]{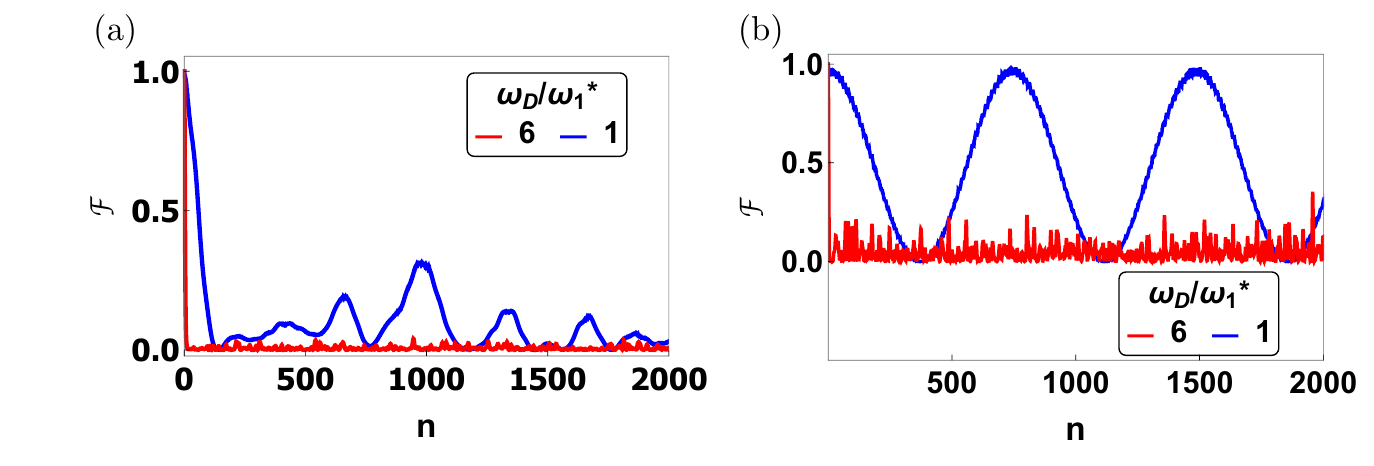}
\caption{(a) Evolution of fidelity $\mathcal{F}$ starting from a
random frozen state for $V_1=19$ and different values of the drive
frequencies. ${\mathcal F}$ remains finite for $n\le 2000$ for
$\omega_D=\omega_1^{\ast}$ (blue curve) indicating violation of ETH
while it vanishes in accordance with prediction of ETH for
$\omega_D= 6 \omega_1^{\ast}$. (b) Same as (a) but now starting from
$|Z_2\rangle$ initial state. The red curves again show thermal
behavior, as expected of an ergodic system, while the blue curve
exhibits persistent slow oscillation as in Fig.\ \ref{fig3}(c) of
the main text. For both the figures, $V_0=10V_2=2$, $L=14$ and all
energies are scaled in units of $J$.} \label{figs2}
\end{figure}

In this section, we study the evolution of the fidelity
$\mathcal{F}=|\langle \psi(0)|\psi(nT)\rangle|^2$ starting from a
random frozen state and a $|Z_2\rangle$ state, using a square
pulse protocol. We find that their behaviors are analogous to those
observed for equal time local correlation functions as shown in
Fig.\ \ref{fig3} of the main text.

These results are shown in Fig.\ \ref{figs2}. Similar to the
behavior of the density-density correlation function discussed in
the main text, we find that for a random state $\mathcal{F}$ does
not decay to zero, as expected for an ergodic system, for
$\omega_D=\omega_1^{\ast}$ (blue curve); in contrast, for $\omega_D
= 6 \omega_1^{\ast}$, $\mathcal{F}$ vanishes within the first few
cycles (red curve). Moreover, we find a much slower departure of
$\mathcal{F}$ from unity at $\omega_1^{\ast}$. These features
indicate a clear violation of ETH at the special frequencies.

A similar computation starting from the initial $|Z_2\rangle$ state
leads to slow oscillations of $\mathcal{F}$ for
$\omega_D=\omega_1^{\ast}$ and its rapid, ETH predicted, decay at
$\omega_D= 6 \omega_1^{\ast}$ (red curve). The oscillations of
$\mathcal{F}$ can be explained in a similar manner as done for the
correlation function $\chi_1$ in the main text. The bonding and the
antibonding states $|\psi_B\rangle=(|Z_2\rangle + |{\bar
Z}_2\rangle)/\sqrt{2}$ and
$|\psi_A\rangle=(|Z_2\rangle-|\Bar{Z}_2\rangle)/\sqrt{2}$ form two
approximate eigenstates of the exact Floquet Hamiltonian. The
off-diagonal matrix elements connecting these two states receive
contribution from higher order terms in $H_F$ due to tunneling to
states with $N_d \ne 0$. Following a similar argument detailed in
the main text, it is easy to show that if the energy split between
these two states is $2\hbar\alpha_d$, then
$\mathcal{F}\sim\cos^2(\alpha_d nT)$. We note that such coherent
oscillations can only happen if the system is non-ergodic and hence
constitute a signature of fragmentation; however, in addition, such
oscillations require $Z_2$ symmetry breaking of the initial state
and thus does not occur for random frozen initial states.

\section{Continuous drive protocol}
\label{contresults}

In this section, we present results for time evolution of
entanglement entropy $S(nT)$, fermion density-density autocorrelator
$C_L(nT)$, and equal time density-density correlation function
$\chi_1(nT)$ as a function of $n$ for the protocol
$V(t)=V_1\cos{\omega_D t}$.
\begin{figure*}
\includegraphics[width=0.9\linewidth]{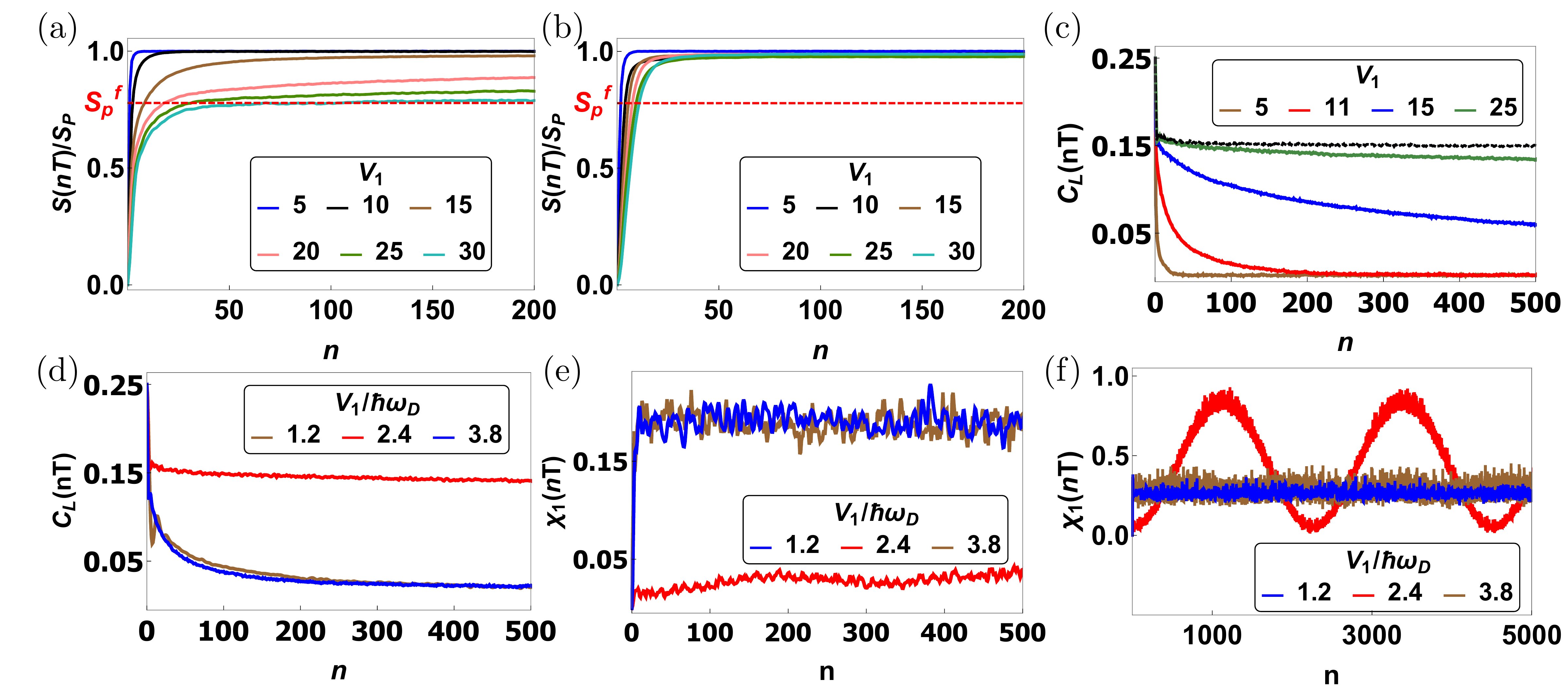}
\caption{(a) Evolution of entanglement entropy $S(nT)$ with drive
cycle $n$ for $V_1/(\hbar\omega_D) = \zeta_1 =2.408$ ({\it i.e.}
$\omega_D= \omega_1^{\ast}$) for different $V_1$ and starting from a
random Fock state. The red dotted line corresponds to the Page value
$S_p^f/S_p$ for the largest fragment of $H_F^{(1)}$ to which the
initial state belongs.(b) Same as (a) but now with
$V_1/(\hbar\omega_D)=\zeta_1/2$. (c) Plot of $C_L(nT)$ as a function
of $n$ starting from a random infinite temperature thermal state for
$V_1/(\hbar\omega_D) = \zeta_1$ and several $V_1$. The black dashed
line corresponds to the value $C_L(nT)$ attains when the evolution
is carried out using $H_F^{(1)}$, while the others correspond to
evolution with exact $H_F$. (d) Same as (c) but now with $V_1=30$
and different frequencies. (e) Plot of $\chi_1(nT)$ as a function of
$n$ for $V_1=30$ and several frequencies starting from a random
frozen state. (f) Same as (e) but now starting from $|Z_2\rangle$.
In (d), (e) and (f), the red curves correspond to $\omega_D=
\omega_1^{\ast}$. For all the plots $V_0=V_2=2$ and all energies are
scaled with $J$.} \label{figs3}
\end{figure*}

From the forms of $H_F^{(1,c)}$ and $H_F^{(1,s)}$ presented in Eq.\
\ref{heff1c} and \ref{hf1s}, it is evident that the former exhibits
the same structure at the special frequencies characterized by
$\omega_m^* = V_1/(\hbar\zeta_m)$ as the latter at frequency
$\omega_m^* = V_1/(2\hbar m)$. Here $m$ is an integer and $\zeta_m$
is the $m^{\rm th}$ zero of $J_0$ as already outlined in the main
text. Thus, we expect to witness similar signatures of prethermal
fragmentation when the system is driven following a continuous drive
protocol.

The plots of $S(nT)$, $C_L(nT)$ and $\chi_1(nT)$ shown in Fig.\
\ref{figs3} confirm this expectation. The results are qualitatively
similar to those obtained for square pulse protocol discussed in the
main text. However, for a given drive amplitude, one witnesses more
stability of HSF for the square pulse protocol as compared to its
continuous counterpart. This is expected since the first non-trivial
correction is third order for square pulse, whereas it is
second-order for the continuous protocol. However, we note that the
qualitative similarity of the plots for the two protocols indicate
protocol independence for realization of prethermal strong HSF.

\end{document}